\documentclass[prc,twocolumn,superscriptaddress,floatfix,nofootinbib]{revtex4}
\usepackage{amsmath}
\usepackage{graphicx}
\usepackage{subfigure}
\newcommand{\iso}[2]{$^{#1}$#2}

\begin{document}

\title{Systematic and correlated nuclear uncertainties in the
  i-process at the neutron shell closure $N=82$}

\author{M.G. Bertolli}\thanks{NuGrid collaboration, http://www.nugridstars.org} 
\affiliation{Theoretical Division, Los Alamos
  National Laboratory, Los Alamos, NM 87545, USA}

\author{F. Herwig}\thanks{NuGrid collaboration, http://www.nugridstars.org}
\affiliation{Department of Physics \& Astronomy,
  University of Victoria, Victoria, BC V8P5C2, Canada}
\affiliation{Turbulence in Stellar Astrophysics Program, New Mexico
  Consortium, Los Alamos, NM 87544, USA}
\affiliation{Joint Institute for Nuclear Astrophysics, USA}

\author{M. Pignatari}\thanks{NuGrid collaboration, http://www.nugridstars.org}
\affiliation{Department of Physics, University
  of Basel, Klingelbergstrasse 82, CH-4056 Basel, Switzerland}

\author{T. Kawano} \affiliation{Theoretical Division, Los Alamos
  National Laboratory, Los Alamos, NM 87545, USA}
  
\date{\today}

\begin{abstract} Nuclear astrophysics simulations aiming to study the origin of the elements in stars require a multitude of nuclear physics input. Both systematic model dependent and statistically correlated uncertainties need to be considered. An application where realistic uncertainty assessments are especially important is the intermediate neutron capture process or i process: a neutron capture regime with neutron densities intermediate between the slow and rapid processes. Accordingly, the main network flux proceeds on the neutron-rich unstable isotopes up to 4-5 species off the valley of stability. The i process has been clearly identified to be active in post-AGB stars during the Very Late Thermal Pulse H-ingestion event, and a recent work infers about its important role in early generations of stars. Here we demonstrate the effect of propagating systematic nuclear uncertainties from different theoretical models to final abundances for a region around the 2$^\mathrm{nd}$ peak at $A-Z=80$ for elemental ratio predictions involving Ba, La and Eu in i-process conditions. These elements are used to distinguish different n-capture contributions observed in low-metallicity stars. For the simple 1-zone model adopted here, predictions vary as much as a factor of 22 in on possible observational plane ([La/Eu] vs. [Ba/La]). 
To consider statistically correlated uncertainties, we similarly perform a full nuclear physics uncertainty study within a given Hauser-Feshbach model and demonstrate the role of correlations on the final stellar abundance uncertainties. We show that in i-process conditions the main result of neglecting correlations is to underestimate the impact of nuclear uncertainties on the final nucleosynthesis yields by as much as two orders of magnitude. In the mass region of the neutron shell closure $N=82$ Cs final abundances are the most affected by correlated nuclear uncertainties with an uncertainty of about a factor of about 3.5 compared to a factor of $6\times 10^{-3}$ when uncorrelated nuclear uncertainties are used.  In both cases Te and I final abundances shows a negligible effect.
\end{abstract}
 
\maketitle

\section{Introduction}
Nuclear physics for stellar nucleosynthesis simulations include charged and uncharged nucleon-nucleus and nucleus-nucleus cross sections as well as weak rates.  Already \citet{Caughlan1988} (see also references therein) emphasized the importance of nuclear reaction rates as a crucial aspect of stellar modeling.  Now, several stellar reaction rate libraries exist such as JINA REACLIB~\cite{Cyburt2010}, NACRE~\cite{Angulo1999}, and KaDoNIS~\cite{dillmann:06}. Rates within these libraries are based on experiment, where available, and theory elsewhere, and usually include some information of the uncertainty of the individual rates.

As the fidelity of stellar astrophysics simulations and astronomical observations of stellar abundances has been continuously increasing, new nuclear astrophysics applications in astronomy have emerged. These include, for example, near-field cosmology \citep[e.g][]{bland-hawthorn:00,venn:04}, the discovery and interpretation of the most metal-poor stars in the galaxy \citep{beers:05,sneden:08} and the investigation of the formation and evolution of the first structure and stars in the universe \citep[e.g.][]{heger:02a,frebel:05,Wise:2011ej}.

\begin{figure}
  \includegraphics[width=0.5\textwidth,angle=0]{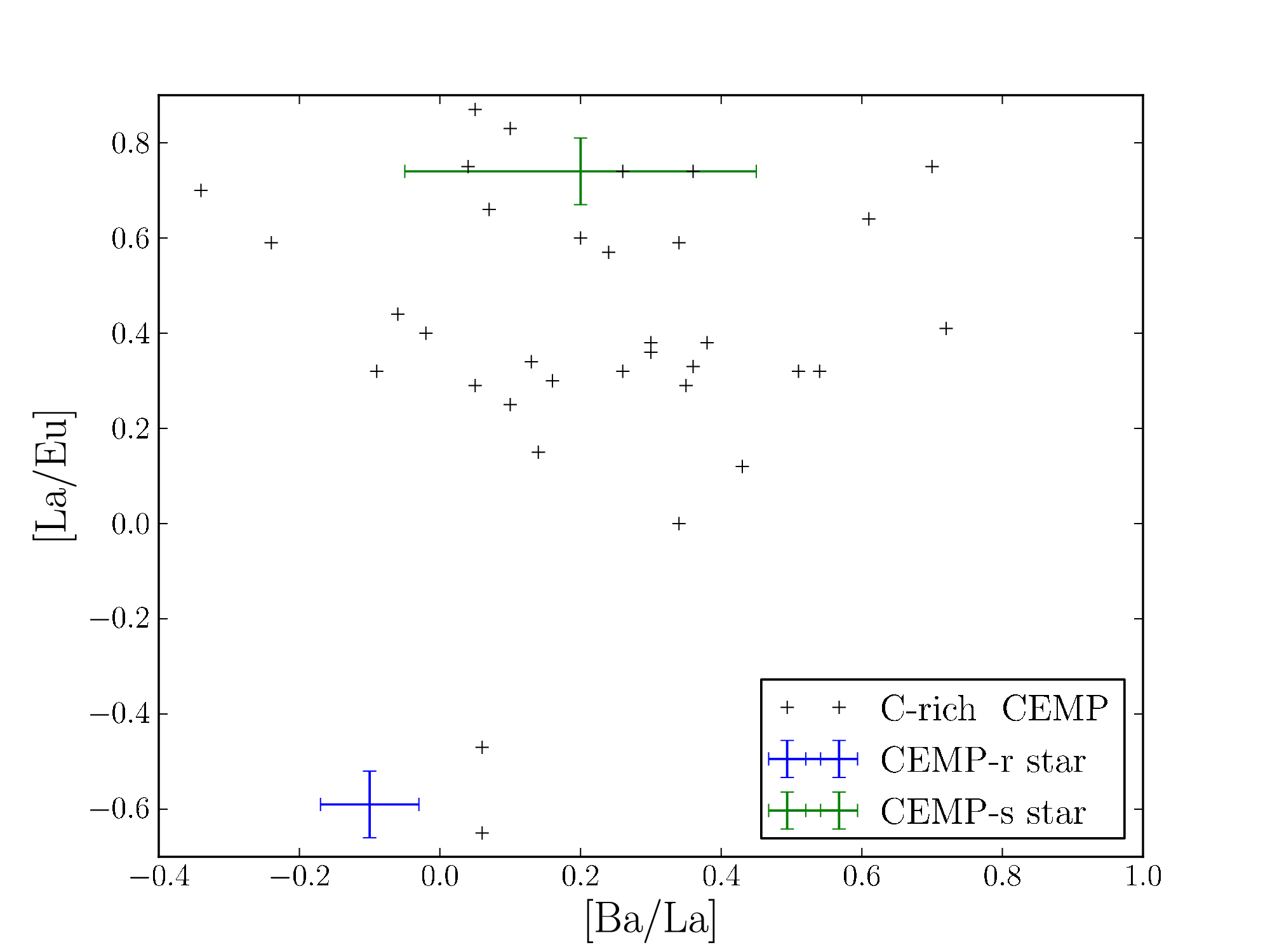}
  \caption{(Colour online) Observed elemental abundance ratios in C-rich and CEMP stars. Data compiled using the SAGA data base \citep{Suda:2008ve}. For one s-process star and one r-process star the typical observational errors are shown. Elemental ratios are shown in bracket notation, i.e. in logarithmic scale normalized to the solar abundances.}
  \label{fig:obs-CEMP}
\end{figure}

Stars of very low metallicity probe the conditions for nuclear production in the early universe. They often show highly non-solar abundance distributions, such as the C-enhancements in a large fraction of metal poor stars (the C-enhanced metal poor or CEMP stars) that are often correlated with strong enhancements of trans-iron elements \citep{beers:05}. Both r-process (such as Eu) and s-process element (such as La and Ba) enhancements are found. Indeed, a number of CEMP stars show an s-process signature close to what s-process simulations predict. For example the CEMP-s stars HD196944 \citep{Aoki:2002fd} has $\mathrm{[La/Eu]} = 0.74$ while s-process models for the same Fe abundance predict a range of $\mathrm{[La/Eu]} = 0.9 \dots 1.0$ (see \citet{Roederer:2010jp}, Table 3). Extensive comparisons are available in \citet{Bisterzo:2012bf} and \citet{lugaro:12}.
The bona fide r-process stars are characterized by large Eu abundance and accordingly $\mathrm{[La/Eu]} = -0.55$ \citep{Sneden:2003gb}. Both stars are shown in a triple-element plot in Fig.~\ref{fig:obs-CEMP} which separates the s- and r-process stars. However, it is clear that most stars lie between the s- and r-process extrema at $0.0 < \mathrm{[La/Eu]} < 0.5$. These stars are usually refereed to as the CEMP-r/s stars, indicating that somehow they carry the signature of both the s and r process. 

Many scenarios have been suggested for the origin of the CEMP-r/s stars. \citet{Bisterzo:2012bf} construct models for a large number of CEMP-s and CEMP-r/s under the assumption that CEMP-r/s stars get the r-process contribution from a pre-enriched cloud out of which a binary forms. The more massive companion will then first evolve to become an AGB star and polute the secondary with the s-process elements. CEMP-s stars would form in the same way, except not out of an enriched cloud. The interpretation of their observational data lead \citet{Allen:2012dc} to the conclusion that Eu must be produced alongside with Ba and C in low-metallicity AGB stars, the primaries of both CEMP-s and CEMP-r/s stars. However, as we pointed out above, this is in stark contrast to present models of n-capture nucleosynthesis in AGB stars, which insist that, e.g. the [La/Eu] ratio must be around one according to s-process models. 

The possible importance of an intermediate neutron-density regime in very metal poor stars was proposed by Herwig \textit{et al.}~\cite{herwig:11}.  \citet{lugaro:12aipc} discuss the possibility that an intermediate process in low-mass AGB stars, triggered by H ingestion, could be responsible for the signature in CEMP-r/s stars. However, this scenario would not explain the systematic larger enrichment of heavy element in CEMP-r/s stars compared to CEMP-s stars, and the preferential correlation with both carbon and nitrogen instead of with only carbon abundance \citep[e.g.,][]{masseron:10}. In this context \citet{herwig:13a} have proposed, based on new stellar evolution models, that a neutron-capture regime with neutron densities intermediate to the slow and rapid process, i.e.\ the i process with $N_\mathrm{n} \approx 10^{15}\mathrm{cm}^{-3}$ \citep{cowan:77}, could be activated in super-AGB stars with Fe metallicities of $\mathrm{[Fe/H]}<-1.3$. These super-AGB stars may be an efficient source of n-capture element enhanced ejecta with i-process signature, which in fact predicts [La/Eu] ratios in the range observed in CEMP-r/s stars.  A related scenario with detailed nucleosynthesis simulations has been explored previously, for example, by \citet{campbell:10} for the He-core flash case, by  \citet{cristallo:09b} for the first convective thermal pulse in low mass AGB stars at low metallicity and by \citet{herwig:11} for the post-AGB star Sakurai's object. Predictions of the latter work can be compared with detailed abundance observations which provide a unique validation opportunity. In addition, the i-process nucleosynthesis predictions for post-AGB stars have been recently shown to correspond to previously unexplained isotopic signatures measured in pre-solar grains \citep{Fujiya:2013ii,2013arXiv1310.2679J}. 

Generally speaking, at the high neutron densities of the i process the network flux proceeds parallel to the valley of stability, offset by approximately $1 - 6$ mass numbers. Here, 
the role of
accurate nuclear physics input becomes a critical issue. Where the experimental
determination of precise cross sections for astrophysically relevant
energies is difficult or impossible,  large uncertainties may be introduced by
theoretical reaction rates, where various models may produce orders of
magnitude differences.  For some time, the detailed evaluation of
nuclear data from both theory and experiment has led to significant
developments in the evaluated nuclear data libraries such as
JENDL-4~\cite{Shibata2011}, JEFF-3.1~\cite{OECD/NEADataBank2009} and
ENDF/B-VII.1~\cite{Chadwick2011}.  Evaluated libraries are produced
comprehensively through critical comparison, selection renormalization
and averaging of the available experimental data with a region and
arecomplemented by nuclear model calculations for specific nuclei
within a region where experimental information may be unavailable.  In
such libraries full uncertainty studies of each reaction may be
included, obtained utilizing various statistical methods.  However,
the evaluated data libraries are often concerned with energy and
isotopic regions not crucial to the astrophysical processes.
Application of these rigorous uncertainty evaluation methods is of
interest to the astrophysics community and has resulted in various
sensitivity investigations of the stellar abundances as a function of
nuclear reaction rate.

In particular, the past decade has produced several studies on the role of neutron capture rates in nucleosynthesis processes, and their effects 
on individual isotopes~\cite{Mumpower2012,Beun2009,Surman2009,Goriely1997,Goriely1998,Rauscher2005,Farouqi2006}.  Sensitivity studies of this sort, considering
only the effects of changes to single reaction rates, are critical for informing which reaction rates are dominant in a given stellar environment.  
They are a first step in providing a complete uncertainty study of the the nuclear physics input.  However, there is no consideration of the systematic uncertainties among 
models or the uncertainty correlations inherent within a model.  These considerations necessitate the inclusion of multiple neutron capture rates in the uncertainty study.  
Such an investigation is needed to determine if single reaction rate uncertainty studies provide sufficient insights to nucleosynthesis uncertainties beyond sensitivities.

In this paper we will demonstrate the propagation of systematic uncertainties among different theoretical models for multiple neutron capture rates in a region of particular importance to the interpretation of CEMP-r/s stars. This is the region around the i-process bottleneck \iso{135}{I} which determines the nucleosynthesis model prediction in the La-Eu-Ba plot (e.g.\ Fig.~\ref{fig:obs-CEMP}) that is particularly suited to distinguish between CEMP-s and CEMP-r/s stars. 

This study begins with a description of the role of systematic uncertainties from varying reaction models.  Additionally, we look at the effects of correlations in determining the full astrophysics uncertainty arising from a single neutron capture model.  We will focus on the effects of theoretical neutron capture rate uncertainties from a Hauser-Feshbach statistical model including model correlations.  The uncertainties within a given Hauser-Feshbach model are developed from physical motivation and propagated using a Monte Carlo method.
	
We investigate the role of large systematic uncertainties and the smaller statistically correlated uncertainties in overall neutron capture rate uncertainties in the i process for 
22 nuclei in the region around the neutron shell closure $N=82$, where $N$ is the total number of neutrons in the nucleus.  The results of propagating uncertainties 
indicate correlations are of significant importance to final abundance uncertainties and astrophysical conclusions.  In Sec.~\ref{sec:rates} we 
will discuss how the neutron capture rates are calculated for a stellar environment and the applicability of the Hauser-Feshbach statistical method.  
Section~\ref{sec:inter-uncert} looks at the proper treatment of uncertainties arising from neutron captures calculated using different Hauser-Feshbach models.  In 
Sec.~\ref{sec:uncertainties} we then consider a Monte Carlo approach to determine the neutron capture uncertainties within a single Hauser-Feshbach model and their 
propagation through the nucleosynthesis simulation.  Finally, in Sec.~\ref{sec:conclusions} we draw conclusions on the importance of correlations within astrophysical 
uncertainty studies.

\section{Nucleosynthesis in the i process}
\label{sec:nucleosynthesis}

In order to analyze the impact of nuclear uncertainties on i-process calculations we use an ad-hoc 1-zone trajectory by Herwig \textit{et al.}~\cite{herwig:13a}.  The 1-zone i-process trajectory is a simplified way of investigating nucleosynthesis at conditions typical 
for the H-\iso{12}{C} combustion in stellar evolution, such as the very-late thermal pulse (VLTP) in post-AGB star Sakurai's object~\cite{Asplund1999,Herwig2011,Werner2006}. In the real 
star these events are characterized by the highly non-linear and non-spherically symmetric interaction of convective flows with violent nuclear burning that demand a full 
3D-hydrodynamic treatment. However, spherically-symmetric multi-zone models can capture important elements of the nucleosynthesis that is encountered in these H-combustion events. 
The key defining property of i-process conditions is the very high neutron density of $N_n \sim 10^{15}~\textrm{cm}^{-3}$~\cite{Cowan1977}. This is a much higher neutron density than the highest s-process values of $N_n \sim 10^{11-12}~\textrm{cm}^{-3}$, which can be found in the He-shell flash convection zone of low-mass AGB stars where neutrons originate from 
the \iso{22}{Ne}-source. But it is also much lower than both n- and r-process conditions with $N_n \sim 10^{18}~\textrm{cm}^{-3}$ and $N_n > 10^{20}~\textrm{cm}^{-3}$ 
(\cite{blake:76} and~\cite{thielemann:11}, respectively)

In real stars i-process conditions can only be obtained in a 2-stage burning and mixing process. First, protons are convectively mixed down into a convectively unstable He-shell 
flash convection zone. The mixing time scale is of the order of $\sim$15 minutes, and the temperature at which the\iso{12}{C}+p reaction time scale is approximately 
$1.1-1.5\times10^8$ K. At this point \iso{13}{N} forms, which is unstable with a half-life of 9.6 minutes. Because it has a higher Coulomb barrier than \iso{13}{C} itÕs 
charged particle reaction cross sections are smaller, and it can be advected by convection to the bottom of the He-shell flash convection zone. Upon arrival and decay the resulting 
\iso{13}{C} will find itself at $T\sim2.8\times10^8$ K and the neutron release happens almost instantly. If ample \iso{13}{N} is produced in the H-burning step then the neutron 
density is defined by the decay rate of \iso{13}{N}, which is model independent. This is the reason we expect nucleosynthesis conditions of a similar type, characterized 
by a typical neutron density, in a wider range of specific objects. This universality justifies identifying these conditions as a separate process, the i process. 

In the past we have studied the i-process conditions in multi-zone nucleosynthesis simulations~\cite{Herwig2011}. However, in order to understand basic nuclear network path 
behaviors a 1-zone trajectory may be more adequate.  We can create i-process conditions in a 1-zone trajectory in a relatively easy way assuming constant 
T and mass density $\rho$. We need to assume an initially high abundance of protons as well as \iso{12}{C} to create copious amounts of \iso{13}{N}. We use a temperature that is 
as high as possible without letting the \iso{13}{N} react with protons to form \iso{14}{O}, which would bypass the \iso{13}{C} neutron source. At the same time we need the 
temperature high enough to release neutrons from the \iso{13}{C}+p reaction as soon as possible after the decay of \iso{13}{N}. A 1-zone burning with constant temperature 
of $T=2\times10^8$K and $\rho=10^4~\textrm{g/cm}^3$, and an initial abundance of X(\iso{1}{H})=0.2, X(\iso{12}{C})=0.5 and X(\iso{16}{O})=0.3485 while all other species are 
according to the solar abundance distribution~\cite{Asplund2005} accomplishes these constraints and provides a neutron density around $N_n\sim10^{15}~\textrm{cm}^{-3}$ 
that is the defining criterion for the i process.

The i-process 1-zone trajectory allows us to investigate general features of nucleosynthesis at such neutron density conditions. Just as the s process and the r process show 
characteristic properties due to the different number of unstable species involved in both processes, we can expect similarly specific signatures of i-process conditions, which are 
the result of a network flux departing much more significantly to shorter half lives away from the valley of stability than the s process, but not reaching the extremely unstable species 
like the r process.  

It is well known that observed signatures of neither r process nor s process can be explained by a single exposure~\cite{Gallino1998,Kratz2008}. The same would have to be true for 
the i process. In the same way as the nucleosynthesis calculation along a single trajectory of a \iso{13}{C} pocket evolution extracted from a complete AGB stellar evolution model 
will give some idea of the general type of nucleosynthesis we can not hope to obtain accurate predictions of the details of a particular observation.

We expect the i-process 1-zone trajectory to provide some general idea of the type of nucleosynthesis that can be expected for environments in which we expect the 
i process to operate, for example in the H-\iso{12}{C} combustion regime in VLTP post-AGB stars, such as Sakurai's object, or the various occasions of this type of events that 
are encountered in the low-Z stars that formed in the early universe. 

The key ingredient of the i process is a combination of rapid proton capture ($p$-capture) nucleosynthesis with rapid neutron capture ($n$-capture) nucleosynthesis. In any 
realistic environment this combination is achieved by a convective connection of these two spatially separated regimes. The i-process 1-zone trajectory combines these 
two regimes in a serial fashion. In the first second all the \iso{12}{C} is burned into \iso{13}{N}. Several other $p$-capture induced reactions take place and overall the network flux 
drives species into the proton-rich side. Around 13.15 minutes, \iso{13}{N} decay starts to provide \iso{13}{C} and an increasing effect of the $n$-captures can be seen, as the flux is 
first drawing the abundance distribution back to the valley of stablility and then driving the network flux into the neutron-rich side.

\begin{figure}[h!]
\begin{center}
\subfigure[Isotopic chart for the first neutron magic peak, $N=50$, at $t = 1.9\times10^{-4}$ years. ]{
\includegraphics[width=0.5\textwidth,angle=0]{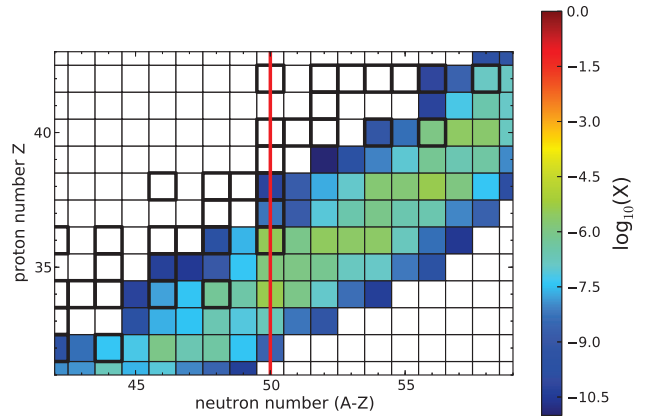}}
\subfigure[Isotopic chart for the second neutron magic peak, $N=82$, at $t = 5\times10^{-4}$ years.]{
\includegraphics[width=0.5\textwidth,angle=0]{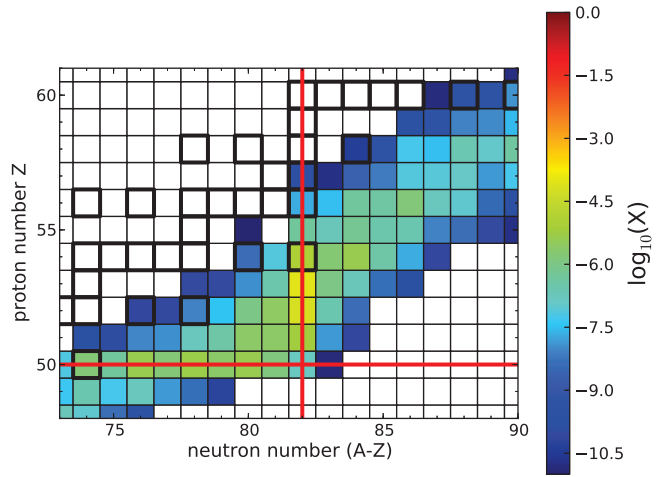}}
\caption{\label{fig:iprocess-flow} (Colour online) The abundance flow around the two neutron 
magic peaks, at key times in the i process.  Red lines indicate magic numbers in either protons or neutrons 
and bold outlines indicate stable isotopes. The present abundance distribution is given in logarithmic-colour 
scale, according to the right band.}
\end{center}
\end{figure}

Neutrons are captured both by light elements as well as by iron and heavier elements. In particular, the neutron capture on the iron seeds leads quickly to the production of 
trans-iron elements Kr, Rb, Sr, Y, Zr around the first neutron-magic peak at $N = 50$.  It peaks initially and then decreases as the network flux moves on to the 
2$^\mathrm{nd}$ peak around Ba and La at $N=82$. These elements peak 
somewhat later. The abundance distribution of Sakurai's object is characterized by a large enhancement of 1$^\mathrm{st}$-peak elements but no enhancement 
of 2$^\mathrm{nd}$ peak elements. 
Figure~\ref{fig:iprocess-flow} shows the abundance flow at the two peaks for select times during the i process.

\section{Neutron Capture Rates}
\label{sec:rates}

Two primary nuclear physics inputs needed to determine the nucleosynthetic outcome are the neutron capture cross sections and rates.  The neutron 
capture cross section $\sigma_\gamma$ may be determined several ways.  Evaluated data on fission product $(n,\gamma)$ reactions is available 
in the various libraries \cite{Shibata2011,OECD/NEADataBank2009,Chadwick2011}.  Additionally, one may calculate the capture cross section directly 
via the Hauser-Feshbach method~\cite{Hauser1952,Wolfenstein1951}.  Hauser-Feshbach is a statistical method giving the 
cross section of reactions that pass through a large number of compound nucleus states~\cite{Hauser1952,Wolfenstein1951,Bohr1936,Hodgson1987}, and 
is implemented in available codes such as {\tt CoH$_3$}~\cite{Kawano2004}, {\tt TALYS 1.4}~\cite{Koning2008,Koning2011} and {\tt NON-SMOKER}~\cite{Rauscher2000}.

The reaction rate for neutron capture at a given stellar temperature $T$ is given by the averaged product of the cross section $\sigma_\gamma$ and the 
relative velocity $v$ between the neutron and target, obeying Maxwell-Boltzmann statistics.  The Maxwellian averaged cross section (MACS) 
$\langle \sigma_\gamma v \rangle$ is then~\cite{Fowler1974}:

\begin{eqnarray}
\lefteqn{\sigma_{\textrm{Maxw}} = \frac{\langle \sigma_\gamma v \rangle}{v_T}} \nonumber \\
&&= \frac{2}{\sqrt{\pi}(kT)^2}  \int_0^\infty dE~\sigma_\gamma(E) E \exp(-\frac{E}{kT})
\label{eq:MACS}
\end{eqnarray}

\noindent where $v_T = \sqrt{\frac{2kT}{\mu}}$ is the thermal velocity, $E$ is the incident energy of the neutron, $\mu$ is the reduced mass of the system and $k$ 
is Boltzmann's constant.  
The cross section $\sigma_\gamma$ is the critical input and source of greatest uncertainty in determining the MACS.  Section~\ref{sec:uncertainties} details how we 
develop the full uncertainties in nucleosynthetic abundances from underlying uncertainties in a particular Hauser-Feshbach model.

\subsection{Applicability of Hauser-Feshbach}
\label{sec:systematics}

\begin{figure}
  \includegraphics[width=0.5\textwidth,angle=0]{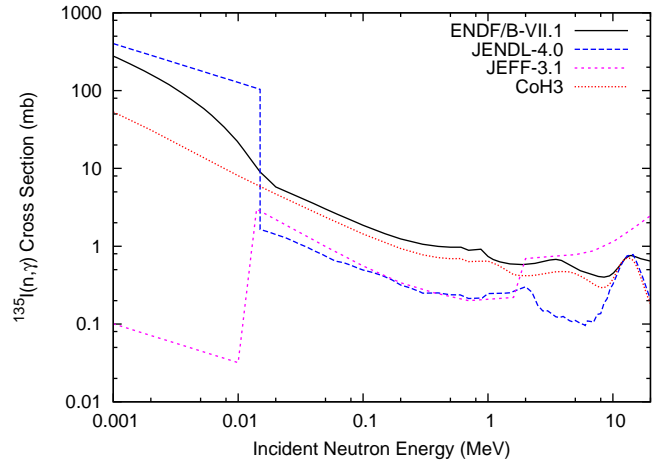}
  \caption{(Colour online) Un-averaged neutron capture cross sections for $^{135}$I as a function 
   of temperature as taken from a statistical {\tt CoH$_3$} calculation and various evaluated libraries.  
   In JENDL-4 and JEFF-3.1 we see the arbitrary connection energy at roughly 10 keV.}
  \label{fig:EvalsvsCoH}
\end{figure}

The Maxwellian temperatures in our astrophysical environment are typically in the keV regions, where the statistical Hauser-Feshbach calculations are 
appropriate. However, the statistical model does not give the $1/v$ cross section at very low energies, since it deals with energy-averaged quantities. In the evaluated nuclear 
data files, there is an energy point that connects the Hauser-Feshbach calculations with the $1/v$ cross section. This connection energy is arbitrary but often determined empirically 
by the average s-wave resonance spacing $D_0$. We have found that the connection energy in the evaluated files has a large impact on the calculated MACS for nuclei close 
to the magic numbers, such as \iso{135}{I}, because relatively large $D_0$ pushes the connection point up in the keV range, causing a drastic change in the evaluated capture 
cross sections. This can be seen in Fig.~\ref{fig:EvalsvsCoH}.

In our study, we do not set the connection energy in order to avoid this unphysical behavior in the calculated MACS. Because the Maxwellian distribution drops quickly at low energies, 
the contribution below 1 keV to MACS is not important. We perform the Hauser-Feshbach calculation from 1 keV, and average the calculated capture cross section over 
the Maxwellian distribution.

\begin{figure}
  \includegraphics[width=0.5\textwidth,angle=0]{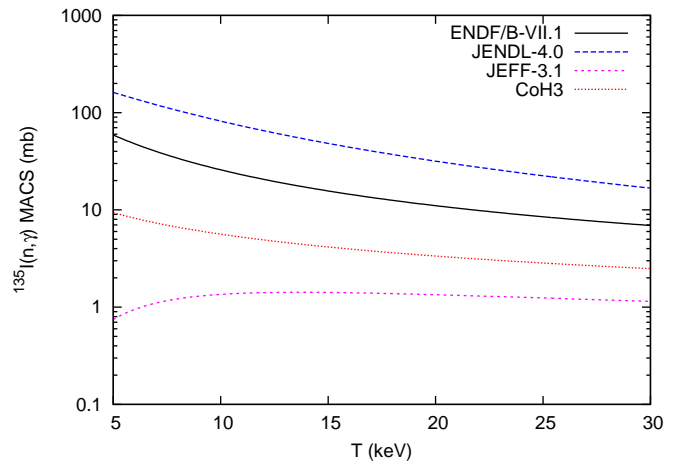}
  \caption{(Colour online) MACS for \iso{135}{I} neutron capture reaction as a function 
   of temperature as taken from the ENDF/B-VII.1 (black), JENDL-4 (blue), JEFF-3.1 (magenta)
    and statistical {\tt CoH$_3$} calculations (red). Calculations begin at an energy of 1 keV, and the temperature range 
   corresponds to energies relevant to astrophysical processes. }
  \label{fig:135IMACS}
\end{figure}

In order to accommodate contributions from the high-energy tail of the Maxwellian distribution {\tt CoH$_3$} calculates the capture cross section for incident energies ranging 
from 1 keV to 10 MeV.  The cross sections are averaged according to Eq.~(\ref{eq:MACS}) in the temperature range of interest, $T=5-30$ keV.  For \iso{135}{I} 
we note that statistical methods are employed in the ENDF/B-VII.1 library at 1 keV.  We show a comparison of the cross sections, non-averaged and averaged, from ENDF 
and {\tt CoH$_3$} in Figs.~\ref{fig:EvalsvsCoH} and~\ref{fig:135IMACS}.  We can see that even when the same general method is applied
there is a relevant difference to take into account.

The Hauser-Feshbach-produced MACS in Fig.~\ref{fig:135IMACS} demonstrates another key aspect of neutron capture cross-sections in astrophysics.  While Hauser-Feshbach is
 applicable in this region, there remain large differences between models due to model parameterizations.  Where no experimental information is available, all cross section 
 calculations rely on some phenomenological inputs.  These differences can lead to as much as a factor of 20 in the calculated MACS in the case of \iso{135}{I}.  A first step in 
 investigating the nuclear  uncertainties in the i process is to consider these inter-model differences as systematic uncertainties in Sec.~\ref{sec:inter-uncert}.

\section{Inter-model Uncertainties}
\label{sec:inter-uncert}

Differences between Hauser-Feshbach models can lead to large uncertainties in the calculated neutron capture MACS.  As a first step in an uncertainty study we consider the trend of 
one model to over- or under-estimate the calculated MACS compared to another model.  In order to see the effect of these model differences on abundances it is first necessary to 
determine the reaction rate, $N_A\langle \sigma_\gamma v \rangle$, from the MACS:

\begin{eqnarray}
N_A\langle \sigma_\gamma v \rangle &=& N_Av_T\sigma_{\textrm{Maxw}} \nonumber \\
&=&N_A\sqrt{\frac{2kT}{\mu}}\sigma_{\textrm{Maxw}}(T),
\label{eq:rate}
\end{eqnarray}

\noindent where $N_A$ is Avogadro's number.  

The nucleosynthetic abundance uncertainties are investigated from the perspective of basic nuclear physics inputs.  Therefore, we start from the point 
of uncertainties generated by the use of two different models for the neutron capture cross section $\sigma_\gamma$ that appears in Eq.~(\ref{eq:MACS}).    For the i process we look 
at the effect of neutron capture rate uncertainties on 22, unstable fission product nuclei in the region of \iso{135}{I}: \iso{132-136}{Te}, \iso{133-137}{I}, \iso{135,137,138}{Xe}, 
\iso{136-139}{Cs}, \iso{139-140}{Ba}, and \iso{137,140,141}{La}.  Limited experimental information is available on the relevant nuclei, and so $\sigma_\gamma$ 
is calculated theoretically with in the framework of statistical Hauser-Feshbach.  The stable isotopes in this region, \iso{134,136}{Xe}, \iso{136-138}{Ba} and \iso{138,139}{La}, as well as the 
very long-lived \iso{135}{Cs}, are neglected in this study because experimental and evaluated data from KADoNIS~\cite{dillmann:06} are adopted by our nucleosynthesis code~\cite{Pignatari:2012dw}.  We consider the KADoNiS rates to be the best choice for a given astrophysical neutron capture rate at 30 keV.  

We will compare the three Hauser-Feshbach codes {\tt CoH$_3$}~\cite{Kawano2004}, {\tt TALYS 1.4}~\cite{Koning2008,Koning2011} and {\tt NON-SMOKER}~\cite{Rauscher2000}.  
{\tt NON-SMOKER} provides the default neutron capture rates for nuclei in the region of interest for these simulations.  It is commonly employed by the astrophysics community through 
its use for theoretical rates in the JINA REACLIB~\cite{Cyburt2010} library.

\begin{figure}[h!]
\begin{center}
\subfigure{
\includegraphics[width=0.5\textwidth,angle=0]{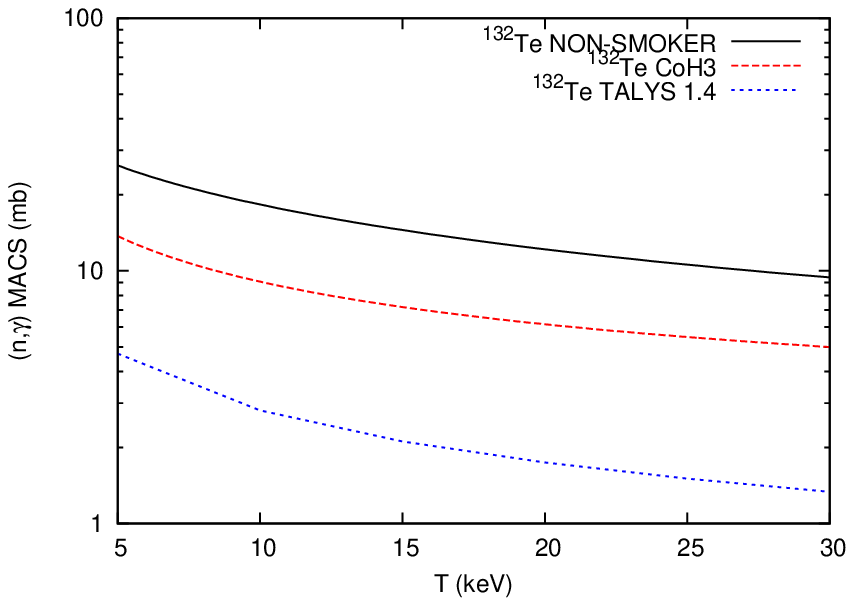}}
\subfigure{
\includegraphics[width=0.5\textwidth,angle=0]{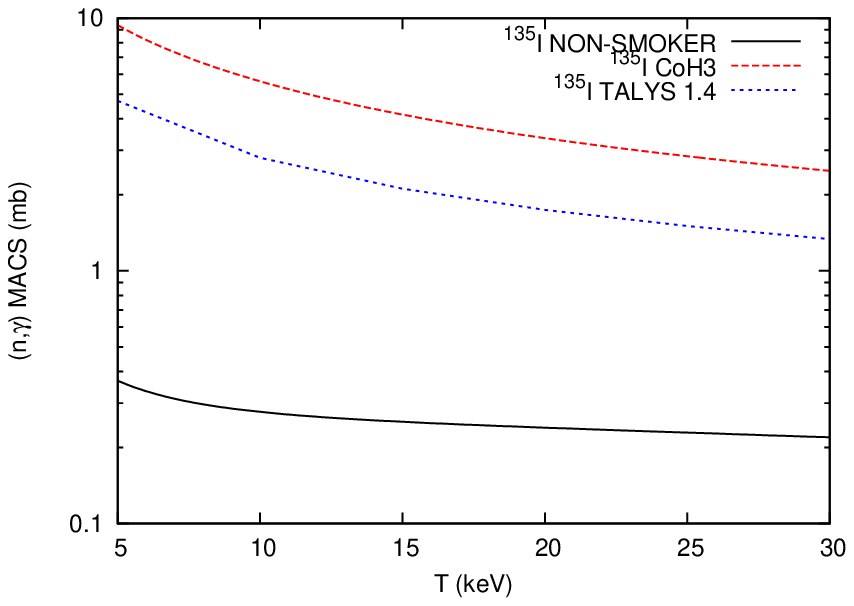}}
\subfigure{
\includegraphics[width=0.5\textwidth,angle=0]{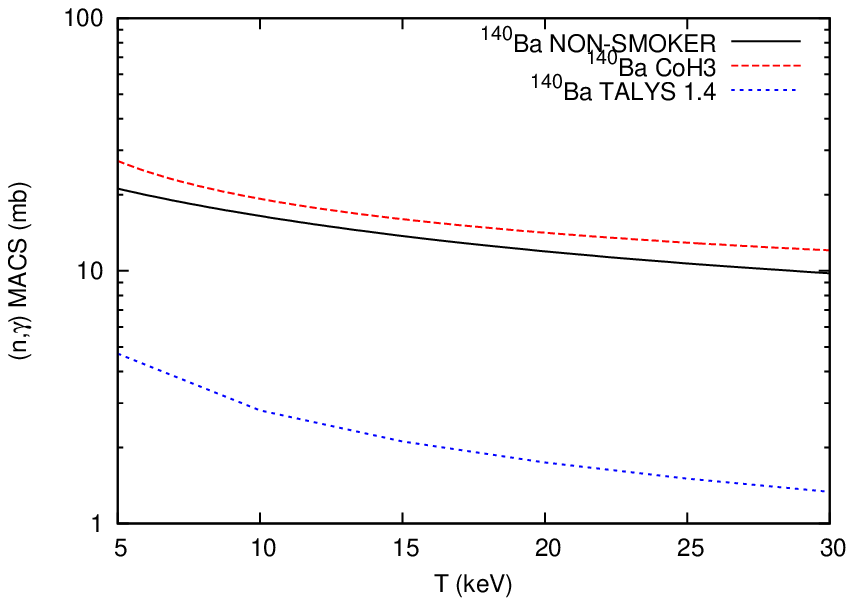}}
\caption{(Colour online) Comparing the neutron capture MACS on a log scale as a function of 
temperature from {\tt NON-SMOKER} (black), {\tt CoH$_3$} (red) and {\tt TALYS 1.4} (blue) 
for three unstable nuclei in the region of interest: \iso{132}{Te}, \iso{135}{I} and \iso{140}{Ba}.  
These nuclei show how the differences among the models cannot be accounted for by a 
single, constant offset.}
\label{fig:CompareMACS}
\end{center}
\end{figure}

The MACS produced by {\tt TALYS 1.4} under-estimates the MACS from {\tt NON-SMOKER} on average by 4\% in this region.  {\tt CoH$_3$}, on the other hand, 
over-estimates the MACS from {\tt NON-SMOKER} on average by a factor of about 1.4.  However, this trend is not always maintained.  Some exceptions are shown in 
Fig.~\ref{fig:CompareMACS}, comparing the neutron capture MACS on a log scale as a function of temperature from the three models for three unstable nuclei in the region of 
interest: \iso{132}{Te}, \iso{135}{I} and \iso{140}{Ba}.  These nuclei show how the differences among the models cannot be accounted for by a single constant offset, despite a general 
trend.  We therefore expect that a systematic application of uncertainty for each of the 22 nuclei in question which accurately reflects the large model differences is necessary to fully 
represent the effect of neutron capture uncertainties on the i-process abundances.

\begin{figure}
  \includegraphics[width=0.5\textwidth,angle=0]{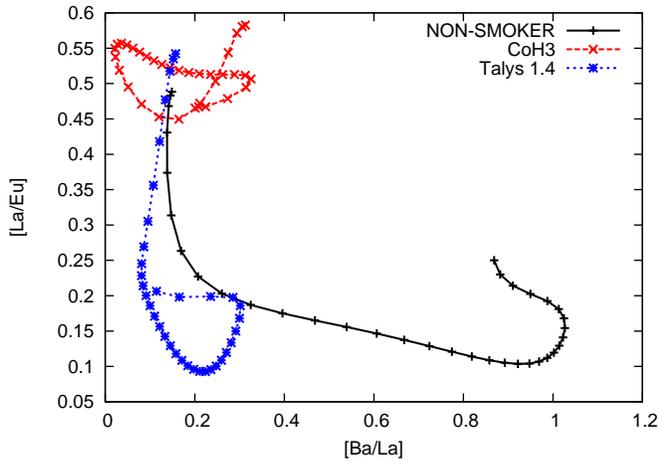}
  \caption{(Colour online) Comparison of the [Ba/La] and [Eu/La] ratios over time when using the default {\tt NON-SMOKER} (black), 
  {\tt CoH$_3$} (red) 
  and {\tt TALYS 1.4} (blue) models.}
  \label{fig:CoHorTALYSNS}
\end{figure}

For each of the 22 nuclei we use the neutron capture rates from the different models to 
investigating the systematic uncertainty amongst them.  The effects of this uncertainty are demonstrated on observable elemental ratios.  
Figure~\ref{fig:CoHorTALYSNS} plots the effects of the uncertainty arising from different models on the [Ba/La] and [La/Eu] ratios. Each point 
represents the ratio of elements at a different time within the simulation, where the 
left most point is the earliest time (about $t = 1.59\times10^{-4}$ years) and proceeding along the curve in increments of about $8\times10^{-6}$ years.   In Fig.~\ref{fig:CoHorTALYSNS}, 
the black curve shows the evolution of these ratios using the default {\tt NON-SMOKER} rates.  The red and blue curves show the element ratio evolution when using the neutron 
capture rates from {\tt CoH$_3$} and {\tt TALYS 1.4}, respectively.

Figure~\ref{fig:CoHorTALYSNS} shows the effect of the systematic differences in neutron capture rates among the three Hauser-Feshbach models.  Between {\tt NON-SMOKER} 
and {\tt CoH$_3$} ({\tt TALYS 1.4}) differences in capture rates can lead to as much as a factor of 22 (47) in the [Ba/La] ratio, and 1.3 (19) in the [La/Eu] ratio.  The differences change not 
only the value of the [Ba/La] and [La/Eu] ratios, but the shape of their evolution.  The evolution of the ratios for the {\tt CoH$_3$} and {\tt TALYS 1.4} rates compared to 
each other and to {\tt NON-SMOKER} demonstrate that the neutron capture rates do not simply cause horizontal or vertical off-sets to the element ratio curves. 

\begin{figure}[h!]
\begin{center}
\subfigure{
\includegraphics[width=0.5\textwidth,angle=0]{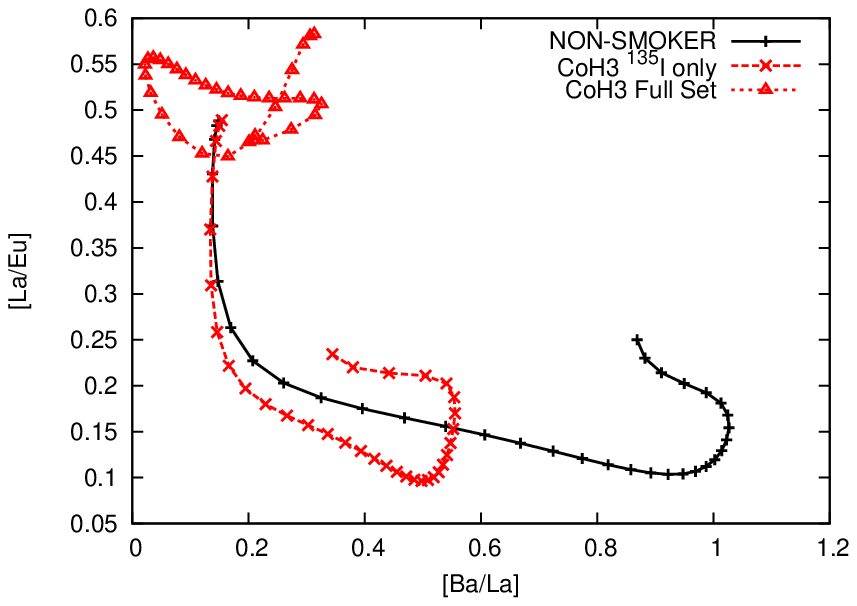}}
\subfigure{
\includegraphics[width=0.5\textwidth,angle=0]{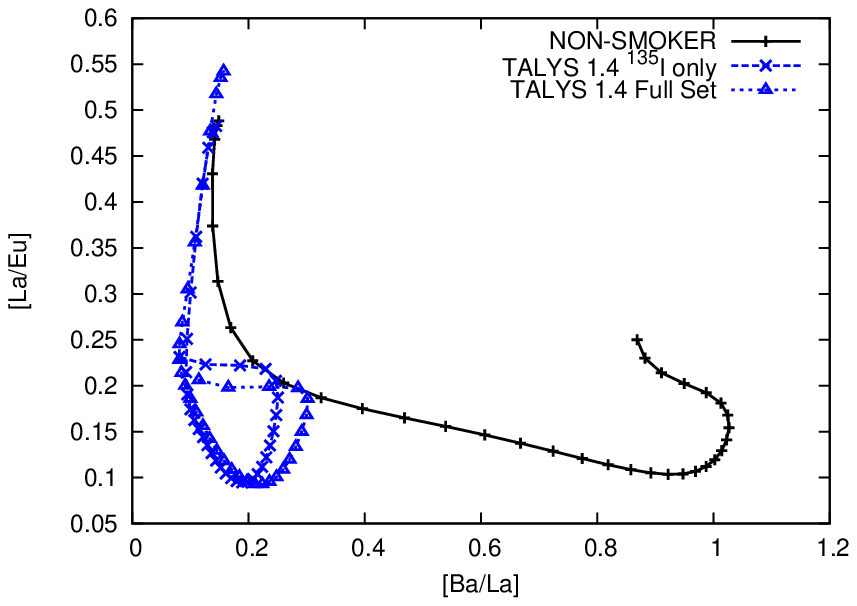}}
\caption{(Colour online)Top: Compares the effects on the [Ba/La] and [La/Eu] ratios when using 
{\tt CoH$_3$} for only the bottle-neck nucleus \iso{135}{I} (red "x") to using {\tt CoH$_3$} for the full set of 22 nuclei (red triangle).  
Bottom: Same as top panel, but using {\tt TALYS 1.4}.  In both cases the elemental ratios from the default {\tt NON-SMOKER} rates are shown in black.}
\label{fig:BNvsFull}
\end{center}
\end{figure}

We are interested in the role of correlated nuclear physics uncertainties on the i-process abundances.  In order to investigate this we look at the application of neutron capture rates from 
different models for all 22 nuclei in the region of the neutron shell closure $N=82$, compared to changing only the bottle-neck neutron capture cross section on \iso{135}{I}.  The 
comparison is shown in Fig.~\ref{fig:BNvsFull}.  While one might expect \iso{135}{I} neutron capture rate to dominate the changes to the [Ba/La] and [La/Eu] ratios, the 
effects of the remaining 21 neighbouring nuclei is important to consider.  
Using different models for only the bottle-neck does change the shape and values of the element ratio evolution for {\tt CoH$_3$} compared to {\tt NON-SMOKER}.  
Yet application of these models to all 22 nuclei results in more drastic changes to the [Ba/La] and [La/Eu] ratios, as well as significant changes to the shape of the evolution.  

On the other hand, the comparison between {\tt NON-SMOKER} and {\tt TALYS 1.4} seems to give a different result. In the lower panel of Fig.~\ref{fig:BNvsFull} there is not
as significant a difference between using {\tt TALYS 1.4} for only neutron capture on \iso{135}{I} than for all 22 
neutron capture rates.  This demonstrates that while {\tt TALYS 1.4} and {\tt NON-SMOKER} neutron capture rates have 
a large difference for \iso{135}{I}, the two codes have similar capture rates for the remaining 21 nuclei in the set.

\begin{figure}
  \includegraphics[width=0.5\textwidth,angle=0]{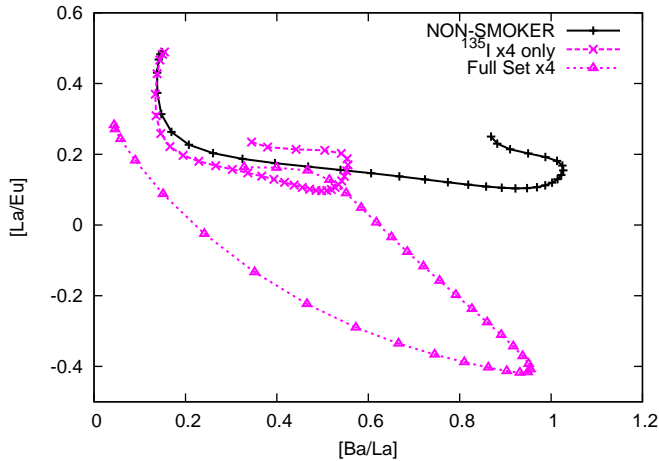}
  \caption{(Colour online) The [Ba/La] and Eu/La ratios over time when applying
  a constant factor of four to only \iso{135}{I} neutron capture (magenta "x") as well as to 
  all 22 nuclei in the region of $N=82$ (magenta triangle).  
  The elemental ratios using the default {\tt NON-SMOKER} rates are shown in black.}
  \label{fig:ConstOffSet}
\end{figure}

However, now consider we apply changes to neutron capture rates in the region but neglect the specific systematics.  We use only an estimated uncertainty of a factor of 
four on neutron capture reactions, which corresponds to the difference between the \iso{135}{I} neutron capture rate in {\tt CoH$_3$} and {\tt NON-SMOKER}.  We plot in 
Fig.~\ref{fig:ConstOffSet} the element ratios resulting from a constant factor of four applied to all 22 nuclei around \iso{135}{I}, as well to only the bottle-neck neutron capture 
on \iso{135}{I}.  Applying the offset to only \iso{135}{I} reproduces a similar behaviour seen in the upper panel of Fig.~\ref{fig:BNvsFull}, as we would expect.  However, the constant factor 
of four results in an even more drastic change to both the [Ba/La] and [La/Eu] values and the shape of the evolution curve when applied to the whole region.  The values of the 
element ratios are significantly reduced and their evolution over time is spread over a larger region.  Thus we see from Fig.~\ref{fig:ConstOffSet} the importance of applying 
neutron capture rate uncertainties with a consideration of systematics, even when considering only these gross, multiplicative off-sets.  Applying uncertainties to only select 
nuclei, or to a region with a constant off-set, will affect the nucleosynthetic abundances in significantly different ways and likely give results inconsistent with underlying nuclear uncertainties.

\section{Systematic and Statistical Uncertainty}
\label{sec:uncertainties}

To consider the effects of detailed statistical correlations on final abundances we now look specifically at a single Hauser-Feshbach model.  We employ the Los 
Alamos National Laboratory (LANL) in-house statistical Hauser-Feshbach code {\tt CoH$_3$} in calculating the $(n,\gamma)$ cross sections 
in the region of $N=82$.  However, within the statistical method large uncertainties still exist.  An analysis of the uncertainty in the calculated capture cross 
sections is performed via a Monte Carlo covariance analysis~\cite{Smith2005}.  The Monte Carlo method simultaneously varies a set of model 
parameters $\mathbf{p}$ around the optimal parameter set.  The optimal parameter set $\mathbf{p}_0$ is determined through a fit to global 
nuclear data.  Each variation of the parameter set is used to generate the capture cross sections, resulting in a history of cross sections at a set of $l$ 
energies.  The statistics of the Monte Carlo histories then generate a global covariance matrix expressing the uncertainty in the calculated cross section via 
Eq.~(\ref{eq:MCCovar})~\cite{Smith2005}.  

\begin{equation}
\mathbf{V}_{ij} =  \begin{array}{ll}
\langle (\sigma_{ik}-\sigma_{i0})(\sigma_{jk}-\sigma_{i0}) \rangle & i,j=1,l
\label{eq:MCCovar}
\end{array}
\end{equation}

\noindent where $k$ is the number of Monte Carlo histories and $k=0$ uses the optimized parameter set $\mathbf{p}_0$.  The uncertainty in the simulation then goes as $1/\sqrt{k}$.

The Monte Carlo used in this work generates the random parameter sets in a manner that accounts for parameter correlations within the model.  
This is achieved by generating a prior correlation matrix in a linearization method via the Jacobian of the model with respect to the relevant 
parameters~\cite{Donaldson1987,Kortelainen2010,Bertolli2011,Bertolli2012}:

\begin{eqnarray}
\mathbf{V}^\textrm{m}_{ij} &\equiv& \chi^2(\mathbf{J}(\mathbf{p}_0)^T\mathbf{J}(\mathbf{p}_0))^{-1} \label{eq:Vm} \\
\mathbf{C}^\textrm{m}_{ij} &\equiv& \frac{\mathbf{V}^\textrm{m}_{ij} }{\sqrt{\mathbf{V}^\textrm{m}_{ii} \mathbf{V}^\textrm{m}_{jj}}} \label{eq:Cm} 
\end{eqnarray}

\noindent where $\mathbf{V}^\textrm{m}_{ij}$ and $\mathbf{C}^\textrm{m}_{ij}$ are the prior model covariance and correlation matrices, respectively.  The covariance matrix in 
Eq.~(\ref{eq:Vm}) depends on the root-mean-square variation, $\chi^2$, that results from fitting the optimal parameter set $\mathbf{p}_0$.  We note that 
because we employ the correlation matrix $\mathbf{C}^\textrm{m}_{ij}$, there is no longer a dependency on $\chi^2$.  Other methods for approximating a model covariance 
exist, however this is the preferred method of Refs.~\cite{Donaldson1987,Kortelainen2010} due to its simplicity, numerical stability and low computational cost.  
Equation~(\ref{eq:Cm}) is then used in the (now correlated) random sampling of parameter sets, by projecting appropriate Gaussian-distributed random numbers 
onto the correlated space.  This allows for a first-order approximation to the prior covariance matrix, rather than assuming a purely diagonal prior covariance as is 
frequently used~\cite{Smith2005,Bertolli2012}.

While the statistical Hauser-Feshbach methods employ a large number of parameters that may be varied, it is sufficient to perform a Monte Carlo covariance 
analysis on a small set of the most crucial parameters.  The set of seven parameters varied in this analysis are related to the optical model, level density 
parameters, and the average $\gamma$-ray width.  {\tt CoH$_3$} uses the global optical model parameterization provided by Koning and Delaroche~\cite{Koning2003}.  
We adjust the real potential depth, $V_V$, and imaginary surface potential depth, $W_D$.  These potential depths have a reported uncertainty of 2\% and 10\%, 
respectively.  We also adjust the real potential radius, $r_V$, and imaginary surface potential radius, $r_D$, within their reported uncertainty of 0.001 fm for both.

In addition to the four optical model parameters listed above, we adjust two parameters of the Gilbert-Cameron level density model~\cite{Gilbert1965}.  The pairing energy, 
$\Delta$, and the level density parameter, $a$, are adjusted within an uncertainties of 30\% and 2.4\%, consistent with systematics reported in Ref.~\cite{Gilbert1965}.  The 
last parameter is the average $\gamma$-ray width, adjusted within an uncertainty of 30\ which was roughly estimated by experimental average $\gamma$-widths, 
$\langle \Gamma_\gamma \rangle$.  All seven parameters are varied simultaneously in the Monte Carlo fashion described above, where random variations are generated 
from a Gaussian distribution with a standard deviation equal to the parameter's reported uncertainty.  Correlations amongst all the parameters within the model are accounted 
for via Eq.~\ref{eq:Cm} and the above described projection.

\begin{figure}
  \includegraphics[width=0.5\textwidth,angle=0]{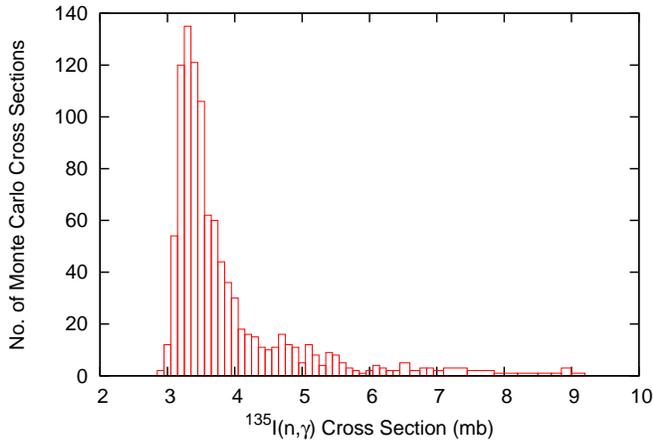}
  \caption{(Colour online) Distribution of Monte Carlo \iso{135}{I} neutron capture cross sections at 30 keV in 0.1 mb bins.  
  The variance of this distribution, given by Eq.~\ref{eq:MCCovar}, provides the uncertainty in the cross section at this 
  energy.}
  \label{fig:135IDistro}
\end{figure}

These parameters are chosen because the model and underlying physics depends sensitively on their value.  Figure~\ref{fig:135IDistro} shows the distribution of 
\iso{135}{I} neutron capture cross sections at 30 keV incident neutron energy which results from the Monte Carlo variation.  The variance of this distribution 
provides the uncertainty in the calculation.  

\begin{figure}
  \includegraphics[width=0.5\textwidth,angle=0]{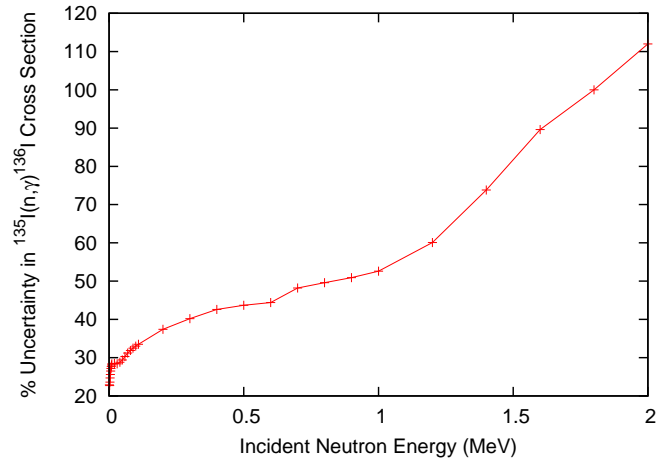}
  \caption{(Colour online) Percent uncertainty in the \iso{135}{I} capture cross section as a function of incident neutron energy. 
  The uncertainty is a monotonically increasing function of energy.}
  \label{fig:135ICovar}
\end{figure}

In Fig.~\ref{fig:135ICovar} we show the percent uncertainty in the {\tt CoH$_3$} calculated \iso{135}{I} capture cross section as a function of incident neutron 
energy from the Monte Carlo covariance analysis.  As one can see, the model uncertainty is not negligible and provides a necessary constraint on the 
uncertainties present in a crucial astrophysical reaction.  The percent uncertainty is shown to go above 100\% at incident energies higher than 1.5 MeV.  This comes from 
the long tail in the asymmetric distribution seen in Fig.~\ref{fig:135IDistro}, where the calculation of the percent uncertainty assumes a Gaussian (symmetric) probability density function 
via Eq.~\ref{eq:MCCovar}.  As pointed out in Ref.~\cite{Rauscher2012}, captures on excited states of the target nucleus must be considered in an astrophysical environment.  
Therefore, sensitivity studies of nucleosynthesis on MACS are not accurately bounded by the uncertainty in experimental results on only ground state targets and it becomes 
necessary to use the model uncertainty as the appropriate lower bound~\cite{Rauscher2012}.
	
\subsection{Propagating Uncertainties}
\label{sec:propagation}

The goal is to propagate the uncertainties in the Hauser-Feshbach parameter set not just to the cross section, but through to the final abundance of the 
i process.  In order to achieve this, these uncertainties must be propagated through the MACS, rates and the nucleosynthesis simulation.  In all steps, 
a stochastic method is employed allowing us to directly include all non-linear effects.

The Monte Carlo propagation of the Hauser-Feshbach parameter uncertainties through the MACS calculation is straightforward.  The capture cross 
sections at $l$ energies generated by the random simulation of model input parameters is used directly within Eq.~(\ref{eq:MACS}) to generate MACS 
for each Monte Carlo history.  Using the same method of Eq.~(\ref{eq:MCCovar}) we can statistically determine the covariance matrix for the MACS at the 
various stellar temperatures,

\begin{equation}
\mathbf{V}^{\textrm{Maxw}}_{ij} =  \begin{array}{ll}
\langle (\sigma_{\textrm{Maxw},k}(T_i)-\sigma_{\textrm{Maxw},0}(T_i))\\
\times(\sigma_{\textrm{Maxw},k}(T_j)-\sigma_{\textrm{Maxw},0}(T_j)) \rangle & i,j=1,\mathcal{M},
\label{eq:MACSCovar}
\end{array}
\end{equation}

\noindent where $\mathcal{M}$ defines the number of temperatures, $T$, considered.  The same can be applied to investigate the covariance matrix amongst various 
isotopes with $\sigma_{\textrm{Maxw},k}(T_i) \rightarrow \sigma_{\textrm{Maxw},k}(N_i,Z_i,T_i)$.

With the stochastic distribution of MACS available it is possible to propagate these uncertainties through the rate and, subsequently, the nucleosynthesis 
simulations.  We choose to consider the neutron capture at 30 keV because it is the maximum applicable energy for stellar reactions in this scenario.  Where the uncertainty is 
a monotonically increasing function of energy (see Fig.~\ref{fig:135ICovar}) this provides the greatest uncertainty for each reaction, giving the least optimistic perspective of 
abundance uncertainties.

\subsection{Correlated vs. Uncorrelated Uncertainties}
\label{sec:corruncorr}

The neutron flow shows a bottle-neck at \iso{135}{I}.  We stochastically propagate uncertainties from neutron capture reactions on 22 nuclei in the region of \iso{135}{I} through the 
stellar neutron capture rate of Eq.~(\ref{eq:rate}) and on to the final abundances of the i process.  We compare the effects of independent (uncorrelated) changes to \iso{135}{I} 
neutron capture and correlated changes to neutron capture rates on all 22 nuclei in the region on final stellar abundances.  Via Eq.~(\ref{eq:MCCovar}) 
we calculate the percent uncertainty in the neutron capture cross section for each of the 22 nuclei in the region of \iso{135}{I}, displayed in Table~\ref{tab:PercUnc}.  

\begin{table}
\centering
\begin{tabular}{|c|c|c|}
\hline
Isotope & \% Unc. $\sigma_\gamma$ (30keV) & \%Unc. $\sigma_{Maxw}$ (30 keV) \\
\hline
\iso{132}{Te} & 27 & 23\\
\iso{133}{Te} & 43 & 38\\
\iso{134}{Te} & 21 & 23\\
\iso{135}{Te} & 45 & 51\\
\iso{136}{Te} & 26 & 24\\
\iso{133}{I} & 22 & 27\\
\iso{134}{I} & 8 & 28\\
\iso{135}{I} & 28 & 37\\
\iso{136}{I} & 31 & 34\\
\iso{137}{I} & 54 & 45\\
\iso{135}{Xe} & 11 & 22\\
\iso{137}{Xe} & 73 & 34\\
\iso{138}{Xe} & 23 & 22\\
\iso{136}{Cs} & 40 & 46\\
\iso{137}{Cs} & 11 & 30\\
\iso{138}{Cs} & 24 & 44\\
\iso{139}{Cs} & 90 & 29\\
\iso{139}{Ba} & 6 & 30\\
\iso{140}{Ba} & 30 & 22\\
\iso{137}{La} & 43 & 40\\
\iso{140}{La} & 45 & 41\\
\iso{141}{La} & 44 & 38\\
\hline
\end{tabular}
\caption{Percent uncertainty in the Hauser-Feshbach neutron capture 
cross section and MACS at 30 keV for 22 unstable nuclei in the region of $N=82$.}
\label{tab:PercUnc}
\end{table}

The neutron capture cross section uncertainties at 30 keV are as low as 6\% in one case (\iso{139}{Ba}), but in general are a factor of two or more.  Because 
the MACS involves a folding over capture cross sections at several energies it is not surprising that the range of uncertainty is much smaller (right-most column 
of Table~\ref{tab:PercUnc}).  Whereas global parameterization may be very well determined within the Hauser-Feshbach model at 30 keV for certain nuclei, 
the Maxwellian average for a range of incidence neutron energies from the statistical method produces similar uncertainties for each isotope.  We note that 
these uncertainties arise from the statistical method with which we calculated the cross sections, and does not signify those uncertainties that are present 
amongst different models.  The global uncertainties amongst models are significantly larger.  However, by constraining our focus to those uncertainties within a single model we are 
able to address the statistical correlations 
amongst neutron capture rates.

In order to quantify the uncertainty in the final abundance, the Monte Carlo history rates are propagated directly in the nucleosynthesis simulations.  
The direct propagation contains all correlations from the model.  In the correlated case Monte Carlo histories are propagated for all 22 nuclei around $N=82$, whereas the uncorrelated 
case only propagates histories for \iso{135}{I}.  We perform the same statistical uncertainty analysis on the abundances as with 
the neutron capture cross sections and MACS.  From all $k$ Monte Carlo histories we construct the covariance matrix similarly to Eqs.~(\ref{eq:MCCovar}) and (\ref{eq:MACSCovar})

\begin{equation}
\mathbf{V}^{\textrm{elem}}_{ij} =  \begin{array}{ll}
\langle (X_{ik}-X_{i0})(X_{jk}-X_{j0}) \rangle & i,j=1,\mathcal{N},
\label{eq:ElCovar}
\end{array}
\end{equation}

\noindent where $\mathcal{N}$ is the total number of elements in the nucleosynthesis simulation and $X_i$ is the mass fraction of the $i^{\textrm{th}}$ element.  We let $k=0$ be the 
default, {\tt NON-SMOKER} rate-based, abundances from NuGrid in order to include the systematic uncertainty present between {\tt CoH$_3$} and {\tt NON-SMOKER} codes.  This 
gives us a picture of the full systematic and statistical uncertainties.  The diagonal elements of the covariance matrix $\mathbf{V}^{\textrm{elem}}_{ii}$ provide the uncertainty 
bounds on each elemental abundance where

\begin{equation}
\mathbf{V}^{\textrm{elem}}_{ii} = (\delta X_i)^2
\end{equation}

\noindent and $\delta X_i$ is the uncertainty in the $i^{\textrm{th}}$ elemental abundance.

\begin{figure}
  \includegraphics[width=0.5\textwidth,angle=0]{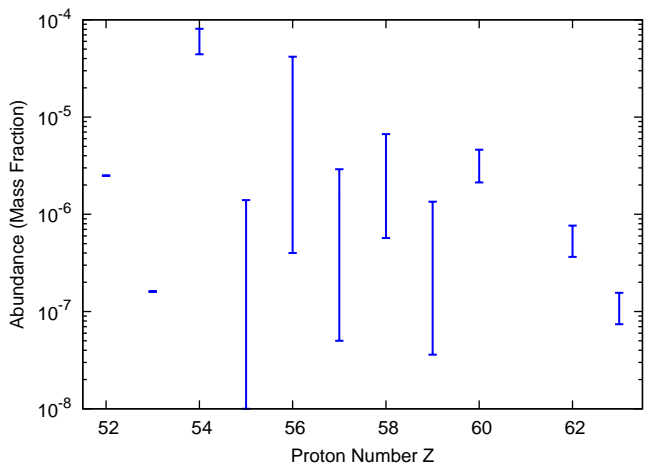}
  \caption{(Colour online) Elemental abundances given as mass fraction of star for the i process.  Error 
  bars indicate uncertainty in elemental abundances from correlated changes to neutron capture rates.  
  Uncertainties are propagated using a Monte Carlo method.}
  \label{fig:CorrErr}
\end{figure}

Figure~\ref{fig:CorrErr} shows the final abundance as mass fraction of the star for relevant elements in the uncertainty study, where all isotopes have been allowed to 
decay to stability.  The error bars indicate the uncertainty resulting from a Monte Carlo propagation of correlated uncertainties in neutron capture rates.  One immediately notices 
the large abundance uncertainties which result from the underlying nuclear physics, even on elements no included in the study (i.e. $Z>57$).  Yet elements Te and I show very 
little uncertainty (factors of $4\times 10^{-3}$ and $6\times 10^{-3}$ uncertainty, respectively), demonstrating the role of coupling to the neutron flow on final abundance uncertainties.

\begin{figure}
  \includegraphics[width=0.5\textwidth,angle=0]{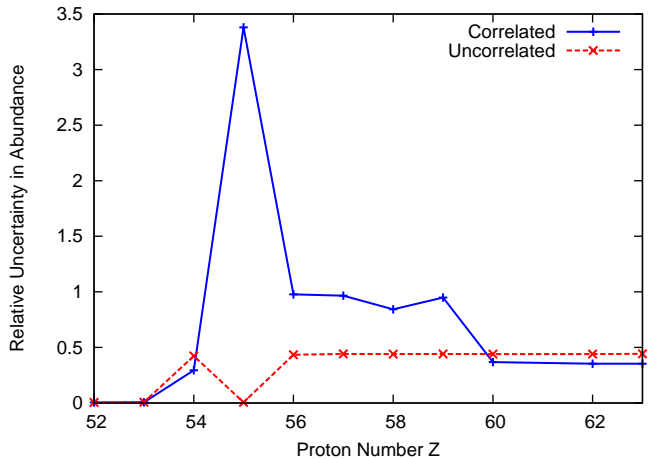}
  \caption{(Colour online) Comparison of fully correlated relative uncertainty for the 22 isotopes 
  studied to uncorrelated uncertainty resulting from changes in \iso{135}{I}$(n,\gamma)$ alone.}
  \label{fig:PercErrProfile}
\end{figure}

Figure~\ref{fig:PercErrProfile} provides an initial insight into the importance of correlated uncertainties for elemental abundances.  The relative error profile in 
abundances due to uncorrelated changes to the \iso{135}{I} capture rate shows a peak at Xe and then remains constant for elements larger than Ba.  These uncertainties 
are a factor of two, consistent with expectations on uncertainties in Hauser-Feshbach models.  However, we see unrealistically low uncertainties for Te, I and Cs 
(factors of $5\times 10^{-3}$, $7\times 10^{-3}$ and $6 \times 10^{-3}$, respectively).  

Correlated neutron capture 
uncertainties result in greater uncertainties in abundances on average.  Additionally, there is more variation in the uncertainty across elements, particularly above 
Ba where the uncorrelated case returns nearly constant uncertainties.  Most notably in the figure, we see a large spike at Cs (factor of about 3.5) for the correlated case, 
resulting from decay of \iso{133}{Te} and \iso{133}{I} whose neutron capture uncertainties are only included in the correlated case.

\subsection{Importance of Statistical Correlation in Nucleosynthesis}
\label{sec:correlation}

The results presented in Figs.~\ref{fig:CorrErr} and \ref{fig:PercErrProfile} show that neutron capture cross section uncertainties can lead to noticeable uncertainties 
in the final elemental abundances.  Including uncertainties from a larger range of isotopes more accurately represents the overall nuclear physics uncertainty and 
allows one to move beyond initial sensitivity studies.  In the case of the i process, correlations in neutron capture uncertainties play a meaningful role in final abundances.  
Figure~\ref{fig:PercErrProfile} has shown a more complex uncertainty is present across elements when correlated uncertainties are used.  Yet the inclusion of correlated 
uncertainties is further demonstrated by considering again the ratio of observable elements. 

\begin{figure}
  \includegraphics[width=0.5\textwidth,angle=0]{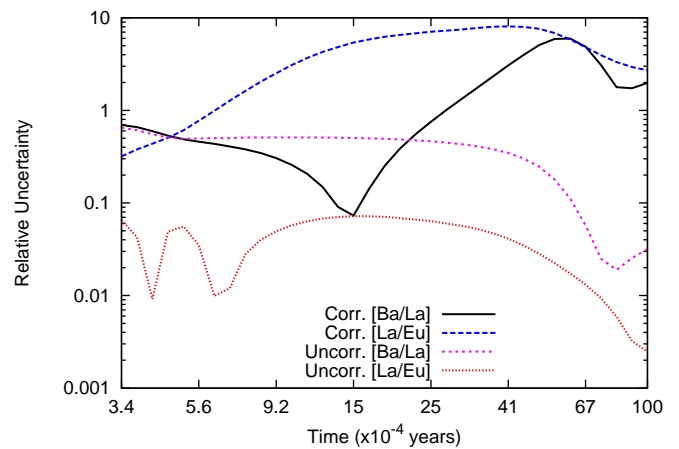}
  \caption{(Colour online) The evolution of the relative uncertainty (on log-scale) in [Ba/La] and [La/Eu] ratios 
  over time resulting from both the correlated and uncorrelated Monte Carlo studies.}
  \label{fig:ratioPerError}
\end{figure}

In Fig.~\ref{fig:ratioPerError} we plot the relative uncertainty of elemental ratios [Ba/La] and [La/Eu] over time, with uncertainties from both correlated and 
uncorrelated propagation of neutron capture rate uncertainties.  The uncertainties shown here are consistent with what we might expect from the differences 
between {\tt CoH$_3$} and {\tt NON-SMOKER} seen in Fig.~\ref{fig:CoHorTALYSNS} when including both systematic and statistical uncertainties.  The 
evolution in the uncertainties in elemental ratios elicits some behaviour of correlated uncertainties in neutron capture rates hinted at in Fig.~\ref{fig:PercErrProfile}.  
Over time the uncertainties change for both ratios, corresponding to variations in the influence of neutron capture on isotope production with the changing stellar 
environment. In the nucleosynthesis calculations done with the 1-zone trajectory 
introduced in Sec.~\ref{sec:nucleosynthesis}, at about $t = 5\times10^{-4}$ years the first neutron flux hits the region of \iso{135}{I}. A major abundance flow continues here in the following trajectory evolution.

 For the [Ba/La] ratio we see that the uncorrelated case is consistently lower than the correlated case.  In earlier times (about $t=3.4-6\times10^{-4}$ years) we see some oscillations 
 in the uncorrelated uncertainty of the [Ba/La] ratio, with the uncertainty becoming as low as a factor of 0.01.  Such a low uncertainty is about a factor of 30 lower than the correlated 
 case, and is  lower than one might reasonably expect.  However, this lower uncertainty is an artifact of the uncorrelated uncertainty study.  As the uncertainties in two quantities 
 approach a correlation of one the relative uncertainty in their ratio approaches zero:  

\begin{equation}
\frac{\textrm{var}(A/B)}{(A/B)^2} = \frac{\textrm{var}(A)}{A^2}+\frac{\textrm{var}(B)}{B^2}-2\frac{\textrm{cov}(A,B)}{AB}.
\label{eq:CancelCovar}
\end{equation}

\noindent In the uncorrelated case \iso{135}{I} acts as a single source of uncertainty causing an artificially high correlation in heavier element production.  More interestingly, the 
oscillations indicate a strong coupling to the stellar environment which influences the level of this artificially high correlation.  Thus, we see the greatest limitation of applying only a 
single rate uncertainty is in describing elemental ratios.  Inaccurately representing the correlation in elemental production with uncorrelated nuclear physics uncertainties result in uncertainties wrongly appearing to cancel, a la Eq.~(\ref{eq:CancelCovar}).

For the [La/Eu] ratio we see a slightly different situation.  The correlated case tends to have a higher uncertainty than the uncorrelated case.  However, we see a decrease of 
about an order of magnitude in the correlated uncertainty at about $t=15\times10^{-4}$ years, below the uncorrelated uncertainty.  During this time, the uncorrelated uncertainty 
remains nearly constant.  This indicates a correlation in the production of La and Eu that occurs in the stellar environment.  However, this production correlation is only captured 
in when correlated changes to neutron capture rates are used in the uncertainty study.  Uncorrelated changes to only \iso{135}{I} do no reproduce this production correlation.  Both 
the correlated and uncorrelated cases have the general shape in the uncertainty curve in the last $20\times10^{-4}$ years, for both the [Ba/La] and [La/Eu] ratios.

\section{Conclusions}
\label{sec:conclusions}

While Hauser-Feshbach is a critical theory for astrophysical neutron capture rates, large differences exist amongst different codes.  Investigating the systematic effects 
on elemental abundances from the codes {\tt NON-SMOKER}, {\tt CoH$_3$} and {\tt TALYS 1.4} has shed light on the role of neutron capture uncertainties in 
the i process beyond single reaction sensitivity tests.  Even with a bottle-neck reaction, such as neutron capture on \iso{135}{I}, the uncertainties present in surrounding 
neutron capture rates play a significant role.  Furthermore, in order to maintain consistency between any two Hauser-Feshbach codes it is not sufficient to apply a single 
offset to all reaction rates indiscriminately.  Rather, one must account for the systematic differences between codes.

Looking beyond systematics, we have provided true error bars on final elemental abundances in the i process by performing a Monte Carlo propagation of nuclear physics 
uncertainties through the stellar simulation, beginning with uncertainties in the parameters of the Hauser-Feshbach code {\tt CoH$_3$}.  By performing our statistical 
comparison directly to abundances using NuGrid's default {\tt NON-SMOKER} rates we are able to describe statistical and systematic uncertainties.

The uncorrelated changes to a single neutron capture rate inaccurately describe the elemental abundance correlations and uncertainties.  We conclude this with the 
assumption that including more physics (i.e. more reaction rate uncertainties) more accurately describes the complete physics of the system.  This is expected, and we 
see from Figs.~\ref{fig:PercErrProfile} and~\ref{fig:ratioPerError} that the role of correlations is significant.  The uncorrelated case tends to under-estimate the uncertainty in 
final elemental abundances, except in early times where we see an artificial oscillation in production correlations.  Elemental abundance ratios are also key observables 
and their uncertainty further indicates the importance of correlations in reaction rate uncertainties. Within a given astrophysical model not including reaction correlations 
leads to an under-estimation of the element ratio uncertainties.  Furthermore, this gives insight into the influence of reaction rate uncertainties and correlations on elemental 
abundance correlations.  

Including uncertainties from 22 neutron capture rates leads to noticeable differences in elemental abundance uncertainties over changes to 
a single reaction rate.  From Fig.~\ref{fig:ratioPerError} we can already see the complex, dynamic environment present in fully propagating nuclear physics uncertainties 
through to elemental abundances.  In particular, we show that in i-process conditions the uncertainty on nucleosynthesis theoretical predictions for the [Ba/La] and [La/Eu] 
ratios may change significantly by using different Hauser-Feshbach models. The sources of these differences need to be analyzed in detail. [Ba/La] and [La/Eu] are indicative observables to 
distinguish for instance between CEMP-s, CEMP-r and CEMP-r/s stars, and therefore provide fundamental insights to analyze the nature of CEMP-r/s stars.   
We have shown that the [La/Eu] is changing by about 0.5 dex by using different Hauser-Feshbach models, while the [Ba/La] by up to 1 dex.
In particular, using the {\tt NON-SMOKER} reaction rates in our simple 1-zone model the i-process results overlap with the observed range of CEMP-r/s stars. In this scenario at least a significant fraction of CEMP-r/s stars are instead CEMP-i stars, according to the conclusions of Herwig \textit{et al.}~\cite{herwig:13a}, and the primary star responsible for the carbon and i-process enrichment might be a super-AGB companion.  Using {\tt TALYS} rates, i-process simulations covered the observed range of [La/Eu] but does not cover the entire range of [Ba/La].
Finally, using {\tt CoH$_3$} the i-process signature is quite similar to an s-process signature for these elements, and therefore is not able to explain the CEMP-r/s abundances spread with the simple 1-zone trajectory.  This shows that while robust i-process predictions depend on the details of the hydrodynamics simulations to correctly follow to H ingestion event 
(e.g.,Herwig \textit{et al.}~\cite{herwig:11}, Stancliffe \textit{et al.}~\cite{stancliffe:11}, Woodward \textit{et al.}~\cite{woodward:13}), there are critical nuclear uncertainties that need to be addressed in the neutron capture flow region typical of the i process. 

Including uncertainties from more reaction rate correlations will lead to even further changes to the uncertainty.  There remain open questions on the role of cross-region correlations, as well as including more than neutron capture rates.  In exotic neutron capture environments, competition with beta-decay is likely to introduce crucial correlations.

This work was carried out under the auspices of the NNSA 
of the U.S. DOE at Los Alamos National Laboratory under 
Contract No. DE-AC52-06NA25396.  NuGrid acknowledges 
significant support from NSF grants PHY 02-16783 and 
PHY 09-22648 (Joint Institute for Nuclear Astrophysics, JINA) 
and EU MIRG-CT-2006-046520. The continued work on codes 
and in disseminating data is made possible through NSERC 
Discovery grant (FH, Canada), and an Ambizione grant of the 
SNSF (MP, Switzerland). The collaboration uses the facilities 
of the Canadian Astronomy Data Centre and the Canadian 
Advanced Network for Astronomical Research operated by the 
National Research Council of Canada with the support of the 
Canadian Space Agency, Compute Canada and CANARIE.ca

\bibliography{i-process,astro}

\end{document}